\documentclass[10pt, aps,prc,twocolumn,superscriptaddress,preprintnumbers,
amsmath, 
floatfix,
longbibliography,
nofootinbib
]{revtex4-1}
\usepackage[T1]{fontenc}
\usepackage[utf8x]{inputenc} 
\usepackage{adjustbox}          % Used to adjust figure frames
\usepackage[caption=false]{subfig}
\usepackage{url}
\usepackage{color}
\usepackage{float}
\usepackage[pdftex,colorlinks=true, linkcolor = blue, citecolor=blue,urlcolor=blue, bookmarksnumbered=true, bookmarksopen=true]{hyperref}
\usepackage{longtable}
\usepackage{amsfonts}
\usepackage{dsfont}
\usepackage{wrapfig,bm} 
\usepackage[normalem]{ulem}
\usepackage{MnSymbol}

\newcommand{\baa}{\begin{align}}
\newcommand{\eaa}{\end{align}}
\newcommand{\beq}{\begin{equation}}
\newcommand{\eeq}{\end{equation}}
\newcommand{\bea}{\begin{eqnarray}}
\newcommand{\eea}{\end{eqnarray}}

\DeclareMathOperator{\sinc}{sinc}

\begin{document}

\title{Multi-Nucleon Transfer Reactions \\and the Emergence and the Evolution of the Compound Nucleus}
  
\author{Matthew Kafker}
\email{mmkafker.physics@gmail.com}
\affiliation{Department of Physics,  University of Washington, Seattle, Washington 98195--1560, USA}
\affiliation{Cyclotron Institute , Texas A\&M University, College Station, TX 77843 USA}
\author{Aurel Bulgac}
\email{aurel.bulgac.phys@gmail.com}
\affiliation{Department of Physics,  University of Washington, Seattle, Washington 98195--1560, USA}

\date{\today}

\begin{abstract}

We present the first implementation of a novel extension of the Generator Coordinate Method (GCM), 
dubbed the enhanced GCM (eGCM), which is applied to the grazing Multi-Nucleon Transfer (MNT) reaction $^{48}$Ca+$^{208}$Pb  
near the Coulomb barrier. eGCM  incorporates major qualitative differences with either Time-Dependent Hartree-Fock (TDHF) or 
GCM frameworks used until now for nuclear reactions.  We demonstrate that the  eGCM framework describes for the first 
time in a fully quantum microscopic framework the emergence and the time-evolution of Niels Bohr's 1936 conjectured Compound Nucleus (CN).
The thermalization time extracted in eGCM is at least two orders of magnitude larger than the Eigenstate Thermalization Hypothesis (ETH) would predict.
 
\end{abstract} 

\preprint{NT@UW-26-13}

\maketitle  

%%\clearpage
%%\newpage
%%\mbox{~}
%%\clearpage
%%\newpage

Ninety years ago Niels Bohr introduced, based on qualitative arguments and some inferences from experimental results, the concept of a compound 
nucleus (CN)~\cite{Bohr:1936,Bohr:1936a,Bethe:1936,Bethe:1937B}, when the individualities of the two initial reaction partners 
are ``irretrievably lost,'' a concept  which has been widely used in nuclear physics ever since. No microscopic derivation based on the
many-body Schr\"odinger equation was ever convincingly introduced and currently the CN is simply a theoretical conjecture, however, amply 
used in phenomenological approaches and consistent with many experimental results~\cite{Bethe:1936,Bethe:1937B,Jandel:2012,Andersen:2007,Morjean:2008,Back:2020,Toke:1985,Rietz:2013,Hinde:2021,Vandenbosch:1973,Gonnenwein:2014}. 
In the editorial~\cite{Bohr:1936a} to the 
\textcite{Bohr:1936} paper, the CN is pictorially described as a superposition of a very large number of ``simpler'' 
many-body states, which basically form a continuum and the best guess anybody currently has is 
that the coefficients in this expansion are complex random numbers, in agreement with  the Wigner-Dyson surmise in Random Matrix 
Theory~\cite{Mehta:1991}, which has been extensively studied in the case of the nuclear shell-model~\cite{Horoi:1995, Zelevinsky:1996}, 
the relation between classical and quantum chaos in quantum systems with a relatively small number of degrees of 
freedom~\cite{Bohigas:1984,Bohigas:1993,Berry:1991,Haake:2018} and for nuclear reactions between heavy-ion nuclei 
or fission~\cite{Ericson:1960,Ericson:1963,Weidenmuller:2009}, processes which are strongly non-equilibrium time-dependent 
phenomena. The loss of memory conjectured by Niels Bohr~\cite{Bohr:1936,Bohr:1936a} is related
classical Boltzmann equation~\cite{Boltzmann:1872}, Poincar\'e recurrence time~\cite{Poincare:1890}, Eigenstate Thermalization 
Hypothesis (ETH)~\cite{Neumann:1929,Neumann:2010,Goldstein:2010,Berry:1977,Berry:1991, Deutsch:1991,Srednicki:1994,Rigol:2012,Bulgac:2024,Abdurrahman:2026a} and  quantum 
scars~\cite{Heller:1984,Heller:2026}, and strong interference between events with extremely large spatial and/or temporal 
separations or heavy quantum objects~\cite{Brown:1954,Brown:1956,Boal:1990,Baym:1998,Merli:1976,Merli:1976,Eichmann:1993,Aspden:2016}, see End Matter and Ref.~\cite{Bulgac:2024d}. 
The multi-nucleon transfer reactions might appear to be in a somewhat better position theoretically, as a theoretical 
framework exists, the Time-Dependent  Hartree-Fock (TDHF) method, with only the nagging questions of whether TDHF has negligible quantum mean field fluctuations  and is accurate enough.

One of the most fundamental questions in physics is how many elements exist and what is the mechanism of their creation, in the laboratory and in Nature, in stars,   
in the collisions of neutron stars and black holes and in supernova 
collapses~\cite{Thoennessen:2004,Thoennessen:2013,Thoennessen:2024,Thoennessen:2016}. 
Elements from $Z=1$ (Hydrogen) to $Z=92$ (Uranium)  were all discovered on Earth until 1939, with a very small number of exceptions,  as those inferred from 
the Oklo natural reactor in Gabon. The number of isotopes which exist naturally on Earth is about 339, including 251 primordial 
isotopes and 88 radioactive isotopes. Among the rest of the predicted 7000-8000 possible 
isotopes~\cite{Erler:2012,Bulgac:2018}, only about 3,000-3,300 have been created 
so far in many laboratories throughout the World~\cite{labs, Valverde:2026}. While most of these predicted 
isotopes are not naturally stable, their even rather short-time existence as an intermediate stage in the creation 
of other elements and isotopes is crucial in order to predict the abundances in the Solar system in particular. The 
overwhelming  number of nuclei and isotopes not found naturally on Earth have been created in multi-nucleon transfer (MNT) 
heavy-ion reactions and the quest to identify the rest of them is an ongoing process. Unfortunately, 
there's still no microscopic framework accurate enough to replace the many 
existing phenomenological models, which are based on various assumptions and are overwhelmingly 
classical in their character. These approaches have been discussed in recent overviews and in 
Refs.~\cite{Loveland:2013,Hinde:2021,Godbey:2025,Zhang:2024a,Li:2024a,Li:2025,Fasoula:2026}. 
In most cases the typical phenomenological equation for production 
of a heavy reaction product used~\cite{Loveland:2013,Hinde:2021,Godbey:2025} is 
\begin{align}
\sigma_{EVR} = \sum_{J=0}^{J_\text{max}} \sigma_\text{cap}(E_\text{CoM},J)P_\text{CN}(E^*,J)W_\text{sur}(E^*,J),
\end{align}
where the CN formation probability $P_\text{CN}$ is phenomenological with wide variations, and
which is the only quantity discussed in this work.

The most sophisticated microscopic model existing in the literature is the 
time-dependent mean field theory in either its TDHF incarnation or in 
its extension, the time-dependent Density Functional Theory (TDDFT)~\cite{Simenel:2010,Sekizawa:2013,Sekizawa:2016,Simenel:2025,Simenel:2025a,Wu:2026},  
an approach which is based on the adequacy of using mean field, which is an expectation of a one-body operator and thus classical in character. This classical character of the mean field 
approximation can be corrected in a configuration interaction (CI) framework, of which the generator coordinate 
method (GCM) was believed for a long time to be a good solution~\cite{Peierls:1957, Griffin:1957,Reinhard:1983,Reinhard:1987}, with some reservations~\cite{Verriere:2017}, 
although this claim was recently challenged with the introduction of the enhanced GCM (eGCM) framework~\cite{Bulgac:2024d}. 

The GCM~\cite{Griffin:1957} is basically a CI-method based on 
linearly independent non-orthogonal system of ground state static (generalized) Slater determinants with fixed shapes, which were assumed 
to describe  an adiabatically evolving nucleus from the neutron capture to scission in induced fission~\cite{Bohr:1939,Hill:1953,Griffin:1957}.
In the case of heavy-ion reactions however the reaction fragments acquire significant excitation energy, not accounted for in an adiabatic 
evolution as \textcite{Reinhard:1983} noticed, which also happens in induced fission~\cite{Bulgac:2016,Bulgac:2019c,Bulgac:2020, Bender:2020}. 
The corresponding many-body wave functions for the three approaches are defined as  
%%%%%%%%%%%%%%%%%%%%%%%%%%%%%%%%%%%%%%%%%%%%%%%%%%%%%%%%%%%%%%%%%%%%%%
\begin{align}
\Psi_{\rm GCM}(\xi_1\ldots\xi_A) = \sumint_{Q} f(Q) \Phi(\xi_1\ldots\xi_A|Q), \label{eq:GCM-s1}\\
\Psi_{\rm GCM_R}(\xi_1\ldots\xi_A,t)=\sumint_Q f(Q,t)\Phi(\xi_1\ldots\xi_A|Q, t), \label{eq:GCM_R}\\
\Psi_{\rm eGCM}(\xi_1\ldots\xi_A) = \sumint_{Q,\tau}f(Q,\tau)\Phi(\xi_1\ldots\xi_A|Q,\tau).
\end{align}
%%%%%%%%%%%%%%%%%%%%%%%%%%%%%%%%%%%%%%%%%%%%%%%%%%%%%%%%%%%%%%%%%%%%%%
The difference between $\Psi_{\rm GCM}$ and $\Psi_{\rm GCM_R}$ is the time dependence of the Slater determinants in the latter~\cite{Reinhard:1983}.
eGCM~\cite{Bulgac:2024d} on the other hand is a ``bouillabaisse''  of linearly 
independent non-orthogonal time-dependent (generalized) Slater determinants, generated with a large set of initial conditions, 
and mixed at all times $\tau$ along their TDDFT trajectories, and thus in the continuum limit
a path integral~\cite{Feynman:1948} along the TDHF trajectories~\cite{Bulgac:2024d}. 

We will demonstrate here that eGCM is a many-body framework 
which leads to many-body wave functions of extreme complexity, 
often quantified by an entanglement entropy~\cite{Bulgac:2022,Bulgac:2023,Bulgac:2024a,Robin:2026}, 
describing the collision of 
heavy ions in particular, which has the expected spatial symmetry from quantum mechanics. We establish
that a very large number of relevant many-body configurations were ignored in all previous incarnations of GCM. 
This is an argument in line with what one would expect from the Fermi golden rule, namely that the number 
of relevant final states is controlled not only by the magnitude of average transition matrix element at each time step, but more importantly, 
by the number of such accessible final states. 

In any scattering of a plane wave of an unpolarized projectile on an unpolarized target the many-body wave function has an azimuthal 
symmetry around the axis of the impinging beam, which is faithfully reproduced in eGCM as implemented here. 
The components of this plane wave corresponding to 
relatively large impact parameters interact with the target only due to the role of the long range Coulomb repulsion 
between the reaction partners and nucleon transfer is suppressed. Only those impact parameters where the two
reaction fragments come into the range of the nucleon-nucleon interaction might significantly contribute to nucleon transfer. One 
has to distinguish at this point between grazing collisions and collisions at small impact parameters of the 
two reaction fragments, which lead to a large variety of  ``CN-like states''  with life times 
varying by many orders of magnitude, see review by \textcite{Hinde:2021} and the End Matter. These various long lived states 
will not be described in detail at this stage into the eGCM framework, 
but may collectively be referred to as the emerging CN. 
For collision energies below the Coulomb barrier, the nucleon transfer is expected to have an essentially tunneling 
under the barrier character and thus be significantly 
suppressed. \textcite{Simenel:2025}
discuss in their recent work the role of the contact time in $^{40\ldots56}$Ca+$^{176}$Yb collisions at energies just 
above the Coulomb barrier and demonstrate that the formation of reaction fragments heavier than the target 
(which is usually the heavier nucleus) occurs only for relatively short contact times between the reaction fragments. 
These are the kind of reactions we will discuss in this initial implementation of eGCM to nuclear reactions.  
We will describe the collision $^{48}$Ca+$^{208}$Pb at a center-of-mass energy 235 MeV, 
which is slightly above the Coulomb barrier between the two initial partners, in both a time-independent and a time-dependent 
frameworks, as these points of view have different physical interpretations. This reaction has been studied experimentally 
and theoretically~\cite{Prokhorova:2008,Hinde:2021}, where the phenomenon of quasi-fission is dominant. At long contact 
times the $^{256}_{102}{\rm No}_{154}$  non-equilibrated nucleus is formed during central collisions of the two initial 
reaction fragments, experiencing both quasi-fission and fission stages.
One can thus separate to a rather good approximation the time evolution of such a collision 
roughly into a fast and a slow phase, characterized by the contact time between the two reaction partners. \\

%%%%%%%%%%%%%%%%%%%%%%%%%%%%%%%%%%%%%%%%%%%%%%%%%%
\begin{figure}[h] 
\includegraphics[width=0.8\columnwidth]{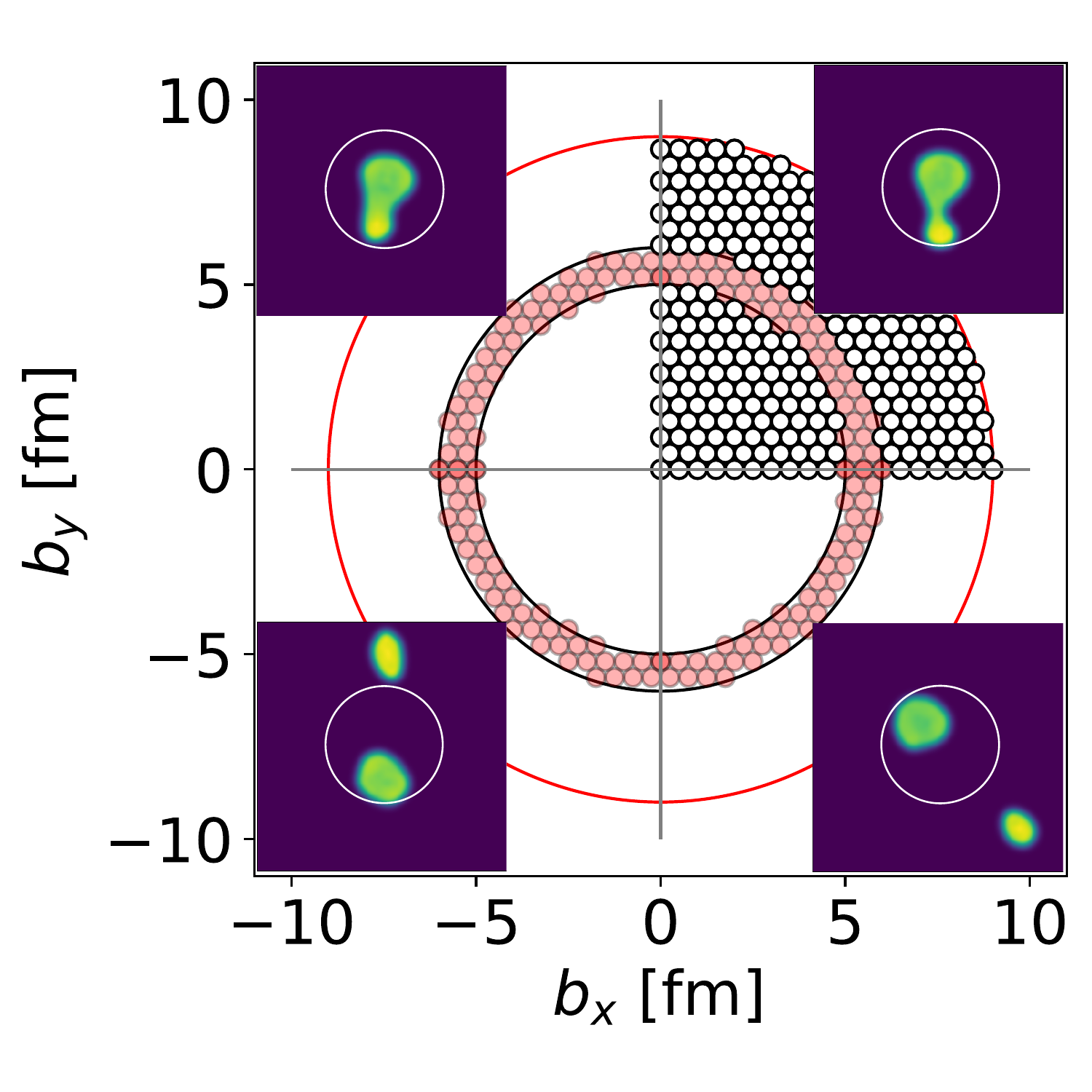} 
\caption{  \label{fig:ring}  
In the impact parameter plane ${\bf b}= ({\rm b}_x,{\rm b}_y)$ only the impact  parameter 
points with centers inside the red ring, with $|\textbf{b}|\approx 5-6$ fm, were
considered. The insets show representative density profiles from the TDHF simulations. On the left are trajectories with ${\bf b}=(5,0)$ and on the right with  ${\bf b}=(6,0)$.
In the upper frames $t=420$ fm/c and the lower frames are shown at their final time. The white circles in the insets indicate the region in which particle number projection is performed at late times.}
\end{figure}  
%%%%%%%%%%%%%%%%%%%%%%%%%%%%%%%%%%%%%%%%%%%%%%%%%%  

As a first step we perform TDDFT simulations (no pairing) of the  collision $^{48}$Ca+$^{208}$Pb at energies slightly above the Coulomb barrier
with the impact parameters within a ring in the impact parameter space shown in Fig.~\ref{fig:ring} and we stored the TDDFT
single-particle wave functions $\phi_k(\alpha)$, where $k=1,\ldots, 256$ and $\alpha=(b_x,b_y,\tau)$, at various times $\tau$ along each TDDFT
trajectory labelled by the impact parameter ${\bf b}$. The TDDFT simulations are performed 
with the energy density functional SeaLL1~\cite{Bulgac:2018} in a simulation box $64^3$ 
with a lattice constant 1 fm and with the code LISE~\cite{Shi:2021}. The storage of the TDDFT single-particle wave functions used 
in our study required about 80 TB of memory, and the construction of the eGCM Hamiltonian matrix was performed using 
48,000 GPUs on the Frontier supercomputer at Oak Ridge National Lab. The number of times $\tau$ the TDDFT 
wave functions stored for use in GCM calculation was $N_\tau =385$ with a step $\Delta \tau = 6.4$ fm/c. We have performed as well some checks of the 
eGCM framework with $\Delta \tau = 3.2$ fm/c.
The total number of impact parameters in the red ring in Fig.~\ref{fig:ring} 
was $N_{\bf b} =152$ and the size of the GCM matrix was 39,630, which is arguably the largest basis set 
used in a nuclear GCM calculation, and particularly in the continuum. For particle number projection in the final state, we used 56,000 GPUs.

Using the eigenvalues and the eigenvectors of the norm overlap matrix between various Slater determinants, 
all linearly independent since $\nu_k>0$, see Fig.~\ref{fig:eGCM_Energies},
we introduce two new sets of orthogonal linear combinations of Slater determinants 
$ |\overline{\Phi}_k\rangle$ and $|\overline{\Phi} (\alpha) \rangle$, in
the formulation of the static eGCM equation, Eqs.~(\ref{eq:GCMS}, \ref{eq:GCMS1})
\begin{align}
&\!\!\!\!\sumint_{\alpha'} \langle\Phi (\alpha) | \Phi (\alpha')\rangle g_k(\alpha') = \nu_kg_k(\alpha ),\, \min(\nu_k) > 0.0011, \label{eq:nu} \\
&\!\!\!\!\sumint_{\alpha} g_k^*(\alpha )g_l(\alpha )=\delta_{kl}, \quad \sumint_k g_k(\alpha)g^*_k(\beta) = \delta(\alpha -\beta), \label{eq:eig}\\
&|\overline{\Phi}_k\rangle =  \sumint_{\alpha}  \nu_k^{-1/2}  g_k(\alpha )|\Phi(\alpha )\rangle, \,  \nu_k>0,
\, \langle \overline{\Phi}_{n} |\overline{\Phi}_{m}\rangle =\delta_{nm}\label{eq:PsiQ} \\
&\!\!\!\! \langle \overline{\Phi}_k | \hat{\rm H} |\Psi_n\rangle =
\sumint_l \langle \overline{\Phi}_k | \hat{\rm H} |\overline{\Phi}_l\rangle \, h_{l n}=h_{k  n}E_n ,\label{eq:GCMS}\\
&\!\!\!\! |\overline{\Phi} (\alpha) \rangle  =\sumint_l  g^*_l(\alpha) |\overline{\Phi}_l\rangle , \quad 
   \langle \overline{\Phi}  (\alpha) | \overline{\Phi} (\beta) \rangle =\delta_{\alpha \beta},\label{eq:Omega1}\\
&\!\!\!\!  \langle \overline{\Phi} (\alpha) | \hat{\rm H} |\Psi_n\rangle =
   \sumint_\beta \langle \overline{\Phi}(\alpha) | \hat{\rm H} |\overline{\Phi}(\beta) \rangle \, h_{\beta n}=h_{\alpha n}E_n, \label{eq:GCMS1}
\end{align}
where $\alpha = ({\bf b},\tau)$, with $\tau$ being the additional and 
crucial generator coordinate that distinguishes eGCM from all previous GCM-type approaches in the literature.
The eGCM formulations in Eqs.~(\ref{eq:GCMS}, \ref{eq:GCMS1}) are each equivalent to a CI solution 
of the many-body problem in the continuum in the present case.
The relation between the many-body wave functions sets  $|\overline{\Phi}_k\rangle$  and $ |\overline{\Phi} (\alpha)\rangle$ is similar to 
the relation between the Bloch and Wannier wave functions in crystals~\cite{Mermin:1976}  
and a similar relation between localized ($\sinc(x)$) and delocalized wave functions in the discrete variable 
representation method (DVR)~\cite{Bulgac:2013}.

If the eigenvectors and the eigenvalues of Eq.~\eqref{eq:GCMS} are determined one can easily construct the solution of 
the time-dependent eGCM (TDeGCM) equation with specified initial conditions
\begin{align}
& i\hbar \partial_t |\Psi(t) \rangle = \hat{\rm H}|\Psi(t)\rangle, \quad \langle \Psi(t)|\Psi(t)\rangle = 1, \label{eq:TeGCM}\\
&|\Psi(t)\rangle = \sumint_{n} \,c_n(t)|\Psi_n\rangle, \, c_{n}(t)= c_{n}(0) \exp\left (-\frac{iE_n t}{\hbar}\right ),\label{eq:psi_t}\\
&c_n(0) 
= \langle \Psi_n|\Psi(0)\rangle ={\cal N}  \sumint_{{\bf b} }\sumint_k g^*_k({\bf b},0)\nu_k^{1/2}h^*_{kn}  ,\label{eq:c_n}
\end{align}
%%%%%%%%%%%%%%%%%%%%%%%%%%%%%%%%%%%%%%%%%%%%%%%%%%
and ${\cal N}$ is the normalization.  The initial many-body wave function  is a uniform superposition 
over all initial states in the ring of impact parameters illustrated in Fig.~\ref{fig:ring}, and so the many-body wave function has by construction the expected azimuthal symmetry. In the TDeGCM with initial conditions as we described above
 the total energy $E_{tot} =\langle \Psi(t)| \hat{\rm H}|\Psi(t) \rangle$ is conserved. Since $|\Psi(t)\rangle$ is a superposition of eigenstates the initial state 
 has a mean energy $E_{tot} = -1817.5$ MeV and a standard deviation $\Delta E= 27.9$ MeV.   Since $\Delta E \Delta \tau < \hbar $ and  initial relative 
 velocity of the two nuclei  $v_{rel} =0.113 c$ is significantly smaller than the Fermi velocity,  the discretization $\Delta \tau$  
 and $\Delta {\bf b}$ (see Fig.~\ref{fig:ring})  appear to be quite satisfactory~\cite{Shannon:1949}.  
 Eq.~\eqref{eq:psi_t} demonstrates that once there is a full solution of any static GCM implementation, there is no need for a  
GCM time-dependent code~\cite{Regnier:2018,Regnier:2016,Verriere:2020,Verriere:2021}.  
One can introduce then the inverse participation ratios (IPR), see End Matter, for the time-dependent many-body wave function $|\Psi(t)\rangle$
\begin{align}
&\!\!\! |\Psi(t)\rangle =   \sumint_k a_k(t)|\overline{\Phi}_k\rangle = \sumint_\alpha  d(\alpha | t) |\overline{\Phi}(\alpha)\rangle, \label{eq:a_alp}
\end{align}
using the either $a_k( t)$ or $d(\alpha | t)$ expansion coefficients.
The probabilities to find $N$ and $Z$ neutrons and protons in the heavy reaction fragment is  given by~\cite{Simenel:2010,Sekizawa:2013,Bulgac:2019a}
\begin{align}
&P_H(N,Z) = \!\!\! \iint\!\!\!\frac{d \eta_N\eta_Z}{(2\pi)^2}\langle  
\Psi (t) |e^{i \eta_N  (\hat{\rm N}_{N} -N) +i \eta_Z  (\hat{\rm N}_{Z} -Z)}|\Psi (t)  \rangle,\nonumber  \\
& \hat{ \rm N}_{N,\, Z} = \sumint_{ \xi }  \Theta_{N,\, Z} ( {\bf r} ) \psi^\dagger( \xi )\psi( \xi) ,\nonumber 
\end{align}
where $\hat{N}_{N,Z}$ is the neutron/proton number operator, $\xi$ stand for space-spin-isospin coordinates, $|\Psi(t)\rangle$ is the final many-body 
wave function  and $ \Theta_{H} ( {\bf r} ) $ is non-vanishing in the heavy reaction fragment 
region.

%%%%%%%%%%%%%%%%%%%%%%%%%%%%%%%%%%%%%%%%%%%%%%%%%%  
\begin{figure}[h]

\includegraphics[width=0.9\columnwidth]{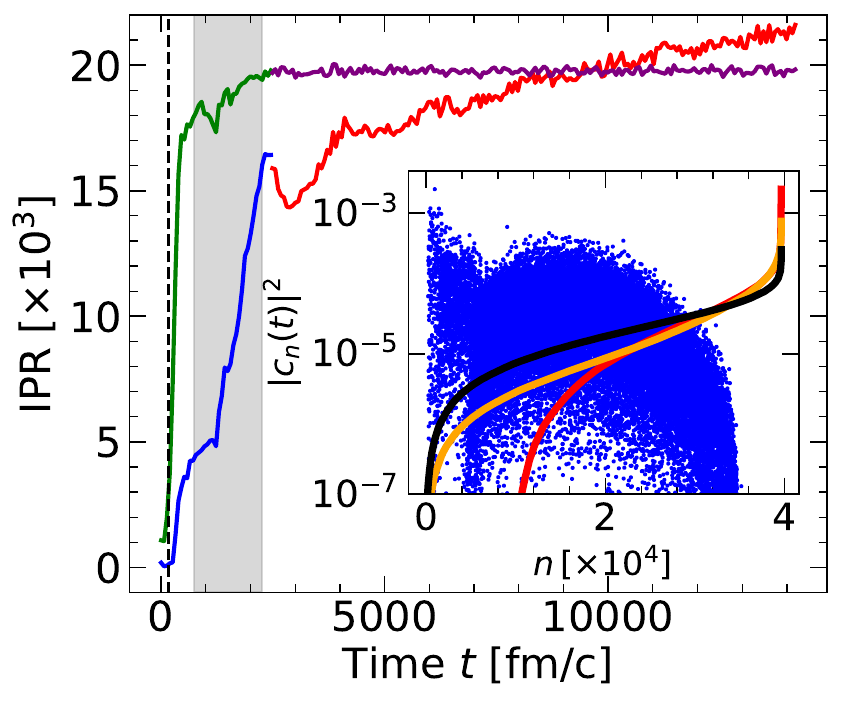}
\caption{ \label{fig:4} 
The IPRs are evaluated for the wave function $|\Psi(t)\rangle$, see Eq.~\eqref{eq:a_alp}, 
using either the expansion coefficients $a_k(t)$ (blue line, continued red after $t_{fin}=2,482$ fm/c), and coefficients 
$d(\alpha|t)$ (green line, continued purple after $t_{fin}$). The vertical black dashed line indicates the time when 
all the nuclei first make contact in TDHF. The gray banded region indicates the range of times during which the nuclei separate in TDHF.
The inset displays the coefficients $|c_n(t)|^2$, as well as the sorted coefficients $|d(\alpha| t_{fin} )|^2$,  
$|a_k( t_{fin} )|^2\times 1.75$, and $|c_n( t_{fin} )|^2$  with black, orange, 
and red  lines respectively. 
}
\end{figure}  
%%%%%%%%%%%%%%%%%%%%%%%%%%%%%%%%%%%%%%%%%%%%%%%%%%  

In the case of the extension of GCM due to \textcite{Reinhard:1983}, denoted here GCM$_R$, both the norm overlap and the 
Hamiltonian overlap matrices have a block-diagonal structure with 
$\langle \Phi({\bf b},\tau )|\Phi({\bf b}',\tau')\rangle \propto \delta_{\tau,\tau'}$ and 
$\langle \Phi({\bf b},\tau) |\hat{\rm H}|\Phi({\bf b}',\tau')\rangle \propto \delta_{\tau,\tau'}$, where the number of 
impact parameters in the present case is typically  $< N_{\rm b}= 152$, 
unlike in the case of eGCM where the dimension of the corresponding matrices is $39,630$ in our simulations 
(arguably the largest GCM simulation ever reported in the literature)~\cite{Bulgac:2024d}. 
The kinetic energies of all the included TDDFT trajectories are 235 MeV, and the total TDHF energy is shown with the thin solid
line in Fig.~\ref{fig:eGCM_Energies} in End Matter. The static eGCM $E_n$ spectrum 
shows an enormous degree of almost degeneracy around the TDHF energy. Within 
eGCM the basis set is constructed from $N_{\bf b}= 152$ full TDHF  trajectories.
For all these impact parameters the reaction fragments separate at the end of the TDHF simulation.  

The inverse participation ratios (IPR) for the norm and energy overlaps provide more insights into the structure and 
the physical interpretations of the eGCM versus the GCM$_R$ wave functions. 
Fig. \ref{fig:3} in End Matter alone demonstrates the overwhelming superiority of the eGCM over GCM$_R$. In case of nuclear 
pairing correlations for example the spreading of nucleon pairs over about ten degenerate orbitals leads to the 
what many would call a ``phase transition'' in a finite system, as seen for example in the magic $^{100}$Sn 
versus the neutron superfluid  $^{120}$Sn.
The eGCM Hamiltonian and norm overlaps  IPRs  are more than two orders of magnitude larger than the maximum value
in the case of GCM$_R$, which is $N_{\bf b}=152$ and which was hoped 
to lead to a satisfactory description of nuclear 
reactions~\cite{Reinhard:1983, Regnier:2019,Hasegawa:2020,Marevic:2023,Marevic:2024,Li:2023,Li:2024,Li:2025}.  

Once the incident projectile and target come into contact, the IPRs start rising dramatically, see Fig.~\ref{fig:4}.
Reaction fragments originating from different impact parameters ${\bf b}$ and ${\bf b}'$  
arrive or leave this interaction region at slightly different times $\tau$ and $\tau'$ and that leads to mixing 
between trajectories after the reaction fragments separates, in the gray region in the Fig.~\ref{fig:4}.

%%%%%%%%%%%%%%%%%%%%%%%%%%%%%%%%%%%%%%%%%%%%%%%%%%  
\begin{figure}[h]
\includegraphics[width=0.9\columnwidth]{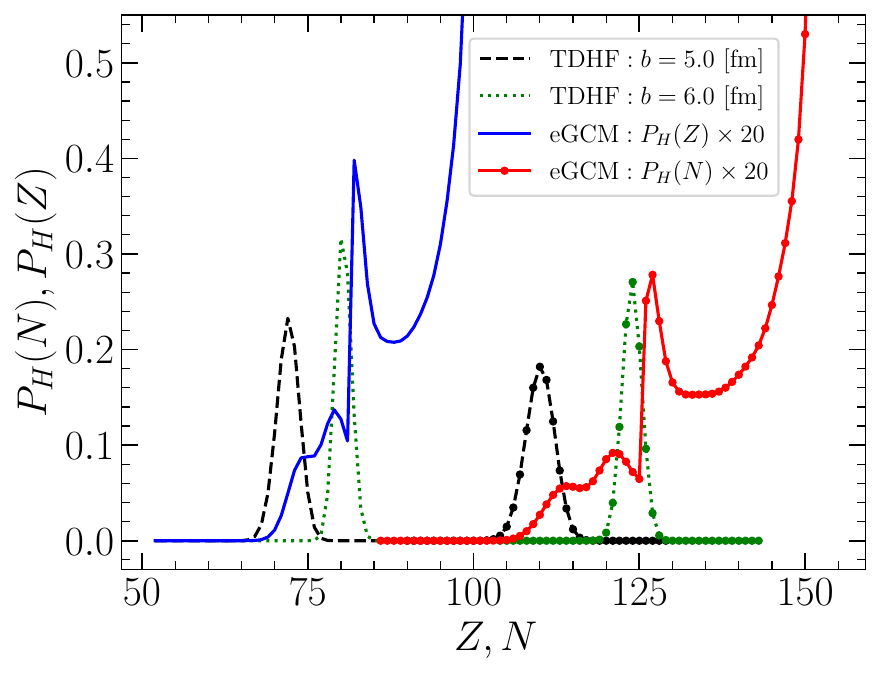}
\caption{ \label{fig:5} Final probabilities for the heavy fragment $P_H(N)=\sum_ZP_H(N,Z)$ and $P_H(Z)=\sum_NP_H(N,Z)$,
 showcasing the dramatic differences between TDHF and eGCM. The black and green lines with additional dots are the TDHF
 probabilities $P_H(N)$, and the black and green lines without addition dots refer to $P_H(Z)$. By contrast, the eGCM distributions $P_H(N)$ and $P_H(Z)$ are shown in red-with-dots and blue, respectively.
 The difference between the TDHF and eGCM particle probabilities are as stark as Isaac Newton's and Thomas Young's 
 predictions would be for the two-slit experiment.  }
\end{figure}  
%%%%%%%%%%%%%%%%%%%%%%%%%%%%%%%%%%%%%%%%%%%%%%%%%%  

A very unexpected result emerging from the eGCM treatment of MNT reactions is the stark difference with 
the results of TDHF simulations. 
In all our TDHF simulations, the two nuclei came into contact in about $t_c=162$ fm/c and all of them 
separated well before we stopped our simulations at $t_{fin}=2,482$ fm/c.  The trajectories 
with $b=5$ fm separate at $t_{sep}=2,262$ fm/c and those with $b=6$ fm separate at $t_{sep}=775$ fm/c, meaning that the nucleus spends $85\%$ and $25\%$ of the total simulation time in the transitional ``CN-like'' system configuration  for $b=5$ fm and $b=6$ fm respectively. 
In eGCM however we observed the emergence of a CN $^{256}_{102}{\rm No}_{154}$, with an excitation energy $E_{ex}\approx 78$ MeV, with $Z=102$ protons and $N=154$ neutrons with a probability $P_H(154,102)=0.34$, which is totally unexpected, 
a result never found in any previous TDHF MNT simulations, see Fig.~\ref{fig:5}. 
The CN does not break apart even when simulated for very long times, a result totally at odds with all TDHF trajectories included in our basis. 
The most surprising outcome however is that $P_H(N,Z)\!\!=\!\!\sum_{N>127}\sum_{Z>83}P(N,Z) \!\!\approx \! 0.85$.
We also find that 
$\sum_{N\leq 126}\sum_{Z> 83}P(N,Z) \!\approx \!0.012$ and $\sum_{N>127}\sum_{Z\leq 82}P(N,Z) \!\approx \!0.006$, indicated that 
MNT favors transfer in either directions of both protons and neutrons, while $P_H(N,Z)\!\!=\!\!\sum_{N\leq 126}\sum_{Z\leq 82}\,P(N,Z)\!\approx \!0.13$.

 We used the eGCM energy spectrum, see  Eqs.~(\ref{eq:GCMS},\ref{eq:GCMS1}), to establish if the properties of our CN are consistent 
with the Wigner-Dyson surmise in Random Matrix Theory~\cite{Mehta:1991,Horoi:1995,Zelevinsky:1996}, 
and thus in agreement with \textcite{Bohr:1936} CN hypothesis, see End Matter. Our eGCM spectrum 
has a quite low average level spacing $\approx 4$  keV, and the normalized distribution of level spacings in eGCM 
has a standard deviation of 0.5273, in close agreement with the value 0.5227 which one obtains in the Gaussian Orthogonal Ensemble (GOE), see also End Matter. 
This demonstrates that eGCM is capable of describing a very complex CN formation.
The proton and neutron numbers, average and standard deviations, for the heavy fragment are  $97.6\pm 7.3$ and $147.7\pm 10.2$, which demonstrates that even in 
grazing collisions a significant number of neutrons and protons are transferred predominantly to the heavy target, in contrast to TDHF, 
see extensive recent published TDHF results~\cite{Simenel:2025}. We have also evaluated the total density of the emerged CN at our simulation and in End Matter we present 
density profiles. The final values of the CN rms radius $\langle r^2\rangle = 7.48$ fm, quadrupole and octpole moments $\langle Q_{20}\rangle =29.23$  b and $\langle Q_{30}\rangle=5.99$ b$^{3/2}$, and 
nucleon density profiles in End Matter at $t=2,482$ fm/c demonstrate that full thermalization and ``memory loss'' have not been achieved.
The CN thermalization time exceeds by orders of magnitude $\tau_{ETH}\approx \tfrac{\hbar}{\Delta E}\approx 7$ fm/c~\cite{Srednicki:1994}, in agreement with thermalization
in other strongly interacting quantum many-body systems Refs.~\cite{Bulgac:2016x,Bulgac:2019,Bulgac:2024}.
The analysis of experimental results~\cite{Prokhorova:2008,Hinde:2021}, is beyond the scope of this first implementation of eGCM.  

eGCM is clearly superior to the tools used so far in the literature, as it emphasizes the major role of the quantum interference and 
of the entanglement among the components of the eGCM many-body wave functions, not encountered so far in nuclear physics. 
The quantum interference is manifested in the high probability to form a CN with a very long stability and  very high $Z$
and $N$, $P(^{256}_{102}{\rm No}_{154})=0.34$, is the quantum expression of Poincar\'e's theorem~\cite{Poincare:1890} of why this emerged CN has an enormous life-time. 
In eGCM the role of nucleon-nucleon collisions, in particular neutron-proton collisions is significantly enhanced. In this work we 
have evaluated only the incipient stages of the CN emergence and evolution.

\vspace{1 cm}

{\it Acknowledgments} \textemdash We thank P. G. Reinhard, I. Stetcu, I. Abdurrahman, G. A. Miller, L. G. Sobotka, T. Kawano, M. J. Savage   
and K. Sekizawa for discussions and comments. AB acknowledges the funding from the Department of 
Energy Office of Science, Grant No. DE-FG02-97ER41014.  This material is additionally based upon work 
supported by the Department of Energy, National Nuclear Security Administration, 
under Award Number DE-NA0004150, the Center for Excellence in Nuclear Training 
And University-based Research (CENTAUR). This research used resources of the Oak Ridge Leadership 
Computing Facility, which is a U.S. DOE Office of Science User Facility supported under Contract No. 
DE-AC05-00OR22725. \\

{\it Data and codes availability} \textemdash The data that support the findings of
this article are not publicly available upon publication,
as it is not technically feasible and/or the cost of
preparing, depositing, and hosting the data would be
prohibitive within the terms of this research project. Reduced 
data sets are available from the authors upon reasonable request. 
Codes, which will become public in time, can be provided without expectations of any guidance.

%
 % These are needed to avoid a babel error.
\providecommand{\selectlanguage}[1]{}
\renewcommand{\selectlanguage}[1]{}

\bibliography{local_fission_3R.bib}
%{\raggedright
%\includepdf[pages={7-8]{gcm+supplement.pdf}
%}

%\bigbreak

%\clearpage
%\newpage
%%\mbox{~}
%%\clearpage
%%\newpage

{\bf  End Matter}\\

{\it Complexity of the Compound Nucleus state.}  \quad An important aspect of Bohr's CN conjecture is the loss of memory of the initial state~\cite{Bohr:1936,Bohr:1936a,Abdurrahman:2026a}.  The loss of memory is similar to Boltzmann's~\cite{Boltzmann:1872} and Poincar\'e's statements about the evolution of  complex mechanical systems, in spite of the fact that they are governed by time-reversal invariant equations.  Schr\"odinger's equation for nuclear systems is time reversal invariant, and thus the statement that memory is lost should be interpreted with care. Naively, one can assume that within TDDFT framework chaos can arise, since the equations are, by construction, non-linear partial differential equations. However, as shown in Refs.~\cite{Shi:2021,Bulgac:2022c}, the Lyapunov exponents in TDDFT are extremely small and practically negligible. Within TDDFT simulations for nuclear or cold atom many-body systems performed with numerical double precision there is no chaotic behavior. In the classical limit, Poincar\'e theorem~\cite{Poincare:1890} relates the loss of memory of the initial conditions to the long Poincare recurrence time, which by transference is expected to apply to a CN. To the authors' knowledge so far no microscopic quantum mechanical framework for the formation and the evolution of a CN was ever suggested. In the editorial~\cite{Bohr:1936a}, published simultaneously with Niels Bohr original paper~\cite{Bohr:1936}, a classical model for the CN was suggested, shown in Fig. 1 in Ref.~\cite{Bohr:1936a}, the collision of a billiard ball with an ensemble of billiard balls at rest in a bowl resting on a table. The impinging ball quickly looses its initial kinetic energy and the ensemble of billiard balls originally in the bowl  can return to rest only if that energy is returned eventually to a single billiard ball, which can then ``escape'' from the bowl. This time however, according to the Poincar\'e's theorem~\cite{Poincare:1890}, is unimaginable long, in agreement with the main assumption of the classical Boltzmann equation~\cite{Boltzmann:1872}, which implies irreversibility, in ``mathematical'' disagreement with the time-reversal invariance of the classical equations of motion. In the limit of quantum many-body quantum dynamics, this topic is vast and related, in part, to the Eigenstate Thermalization Hypothesis (ETH)~\cite{Neumann:1929,Neumann:2010,Goldstein:2010,Berry:1977,Berry:1991, Deutsch:1991,Srednicki:1994,Rigol:2012,Bulgac:2024,Bulgac:2016x,Bulgac:2019} and the present work, quantum scars~\cite{Heller:1984,Heller:2026}, classical chaos versus quantum chaos within the Gaussian Orthogonal Ensemble~\cite{Bohigas:1984,Bohigas:1993,Mehta:1991,Horoi:1995,Zelevinsky:1996,Haake:2018},  where many similarities and disagreements arise when compared to the classical limit. The emerging CN  starts with a relatively simple many-body wave function, see Eq.~\eqref{eq:ini_wf} below, and evolves into a many-body wave function characterized by a complexity which is significantly and quite rapidly increasing in time as demonstrated in the present work, see Fig.~\ref{fig:4}. In eGCM, unlike in all previous implementations of GCM,  a strong interference between different events originating from different either static DFT or TDDFT configurations  with large spatial and/or temporal separations are allowed. This type of interference is well known  in astronomy~\cite{Brown:1954,Brown:1956}, in nuclear physics~\cite{Boal:1990,Baym:1998}, and in two-slit experiments with very low intensity beams, when the particles following different paths hit the screen at very different times~\cite{Merli:1976,Eichmann:1993,Aspden:2016}, or between heavy objects~\cite{Arndt:1999} as can also occur in heavy-ion collisions, as described here, or fission~\cite{Jandel:2012,Andersen:2007,Morjean:2008,Back:2020,Toke:1985,Rietz:2013,Hinde:2021,Vandenbosch:1973,Gonnenwein:2014}, where CN life times are  extracted to be of order of attoseconds.\\

{\it Static and time-dependent eGCM and GOE properties of the eGCM spectrum.}  \quad
Since all the eigenvalues of the overlap matrix Eq.~\eqref{eq:nu} are non-vanishing ($\nu_k \geq 0.0011263$) the entire set of static eGCM Slater determinants $|\Phi(\alpha)\rangle$ 
are linear independent  and the sets of the Slater determinants $|\Phi(\alpha)\rangle$ and of the orthogonal many-body wave functions  
$|\overline{\Phi}_k\rangle$ and $|\overline{\Phi}(\alpha)\rangle$ all have the same size, namely 39,630.  
All many-body wave functions defined in this work depend on $3\times256=768$ spatial coordinates and  256 spin coordinates. 
The static eGCM eigenfunctions $|\Psi_n\rangle$ are defined in Eqs.~(\ref{eq:GCMS}, \ref{eq:GCMS1}) are
\begin{align}
|\Psi_n\rangle = \sumint_l   h_{ l n}  |\overline{\Phi}_l\rangle  =\sumint_\beta h_{\beta n}|\overline{\Phi}(\beta )\rangle.
\end{align}
In GCM one needs to introduce an energy density functional ${\cal E}(n^{\alpha,\alpha'},\tau^{\alpha,\alpha'},\dots) $ depending on the GCM number, kinetic and 
densities~\cite{Ring:2004,Kafker:2025}
\begin{align}
&\hat{M}_{kl}(\alpha,\alpha') = \langle \psi_k(\alpha)|\psi_l(\alpha')\rangle, \\
&\langle \Phi (\alpha) | \Phi (\alpha')\rangle=\text{Det} \,\hat{M}(\alpha,\alpha'). \label{eq:wfs}\\
&n^{\alpha,\alpha'}({\bf r},\sigma |{\bf r}',\sigma')= \frac{\langle \Psi(\alpha)| \psi^\dagger({{\bf r}',\sigma'})\Psi({\bf r},\sigma)
|\Psi(\alpha')\rangle}{\langle \Psi (\alpha) | \Psi (\alpha')\rangle} \nonumber \\
& \quad \quad \quad \quad \quad \quad 
=\sum_{kl} \phi_k(\alpha'|{\bf r},\sigma) \rho^{\bf \alpha,\alpha'}_{kl}\phi_l^*(\alpha|{\bf r}',\sigma') \label{eq:n}, \\
&\rho^{\alpha, \alpha'}_{kl}=\frac{\langle \Psi(\alpha)|c^\dagger_lc_k
|\Psi(\alpha')\rangle}{\langle \Psi (\alpha) | \Psi (\alpha')\rangle}=\hat{M}^{-1}_{kl}(\alpha,\alpha'), \label{eq:over}
 \end{align}
where $\psi^\dagger({\bf r}',\sigma')$ and $\psi({\bf r},\sigma)$ are creation and annihilation field operators 
and $\phi_{k,l}(\alpha|{\bf r},\sigma)$ are single-particle wave functions defining $|\Phi(\alpha)\rangle$.  The GCM energy is~\cite{Ring:2004,Kafker:2025}
\begin{align}
&E =\sumint_{\alpha,\alpha'} \langle \Phi (\alpha ) | \Phi (\alpha' ) \rangle \times  \nonumber \\
& \quad \quad \sumint_{{\bf r},\sigma,\sigma'} {\cal E} \left ( n^{\alpha,\alpha'} ( {\bf r},\sigma | {\bf r},\sigma') ,\tau^{\alpha,\alpha'}({\bf r},\sigma | {\bf r},\sigma') ,\ldots  \right ), \nonumber 
\end{align}
where $\tau^{\alpha,\alpha'}$ is the kinetic energy density and ellipses stand for other  needed densities. 
The notation for spin and isospin degrees of freedom is compounded here as $\sigma$.

%%%%%%%%%%%%%%%%%%%%%%%%%%%%%%%%%%%%%%%%%%%%%%%%%% 
\begin{figure}[h]
\includegraphics[width=0.9\columnwidth]{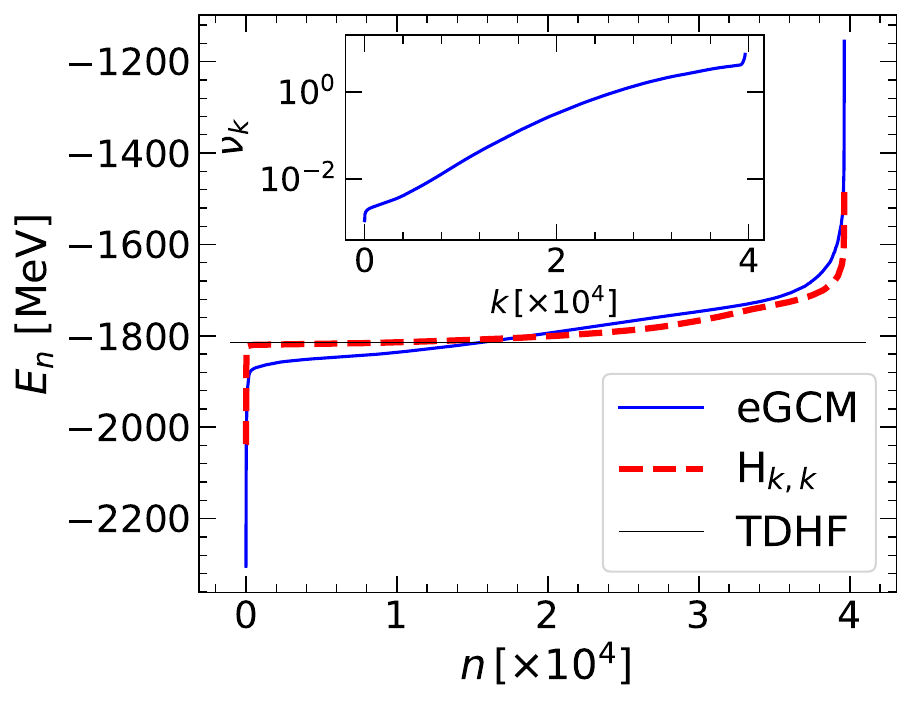}  
\caption{  \label{fig:eGCM_Energies}  
The eGCM energies (blue line) for all trajectories used in this analysis versus the diagonal matrix elements  
$\langle \overline{\Phi}_k|\hat{\rm H}|\overline{\Phi}_k\rangle$ (red dashed line).
The inset displays the eigenvalues $\nu_k$ of the norm overlaps, see Eq.~\eqref{eq:nu}.  }
\end{figure}  
%%%%%%%%%%%%%%%%%%%%%%%%%%%%%%%%%%%%%%%%%%%%%%%%%% 

%%%%%%%%%%%%%%%%%%%%%%%%%%%%%%%%%%%%%%%%%%%%%%%%%%  
\begin{figure}[h]
\includegraphics[width=0.9\columnwidth]{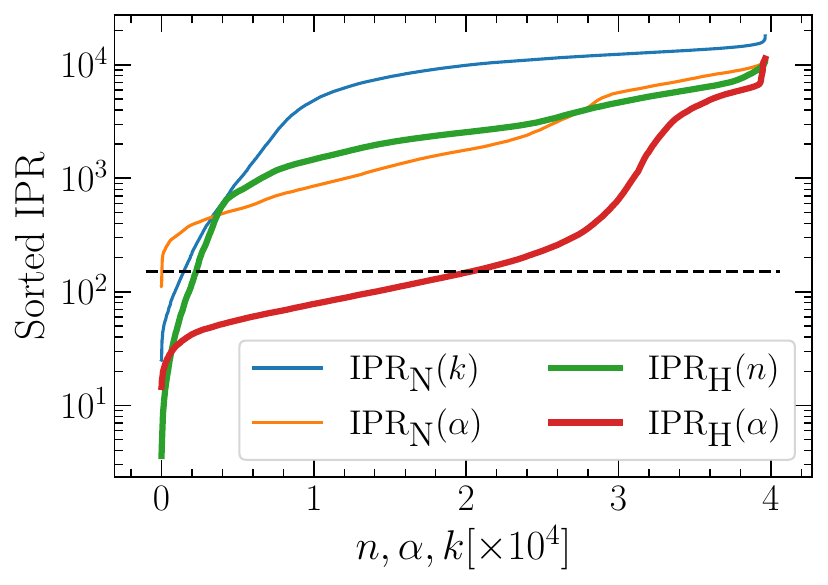}
\caption{\label{fig:3} The IPR for the Norm (thin lines) and the Hamiltonian (thick lines) eGCM overlaps. 
The maximum possible IPR for the GCM$_R$ overlaps is shown with a dashed line. }
\end{figure}  
%%%%%%%%%%%%%%%%%%%%%%%%%%%%%%%%%%%%%%%%%%%%%%%%%  

A very useful measure of the complexity of the eGCM many-body wave functions, which reflects the degree of mixing between 
different  TDDFT trajectories, are the inverse participation ratios (IPR) for the norm and Hamiltonian overlaps of many-body wave functions
in various representations introduced in Eqs.~(\ref{eq:nu}, \ref{eq:PsiQ}, \ref{eq:GCMS}, \ref{eq:psi_t}, \ref{eq:a_alp})
%%%%%%%%%%%%%%%%%%%%%%%%%%%%%%%%%%%%%%%%%%%%%%%%%%% 
\begin{align}
&\!\!\!\! {\rm IPR}_{\rm N} (k)             = \left [ \sumint_{\alpha} |g_k(\alpha)|^4 \right ]^{-1} \!\! ,\, 
            {\rm IPR}_{\rm N} (\alpha)     = \left [ \sumint_{k} |g_k(\alpha)|^4 \right ]^{-1},         \nonumber  \\
&\!\!\! {\rm IPR}_{\rm H} (n)             = \left [\sumint_{k}  | h_{k,n}  |^4 \right ]^{-1}              \!\! ,\quad
           {\rm IPR}_{\rm H} (\alpha)     = \left [\sumint_{n}  | h_{\alpha,n}  |^4 \right ]^{-1},      \nonumber
\end{align}
%%%%%%%%%%%%%%%%%%%%%%%%%%%%%%%%%%%%%%%%%%%%%%%%%%  
see Fig.~\ref{fig:3}. All these many-body wave functions introduced in  Eqs.~(\ref{eq:nu}, \ref{eq:PsiQ}, \ref{eq:GCMS}, \ref{eq:psi_t}, \ref{eq:a_alp}) 
are related by unitary transformations between the many-body basis states, and some are more economical in practice than others 
and in some instances characterized by  a corresponding intrinsic entanglement entropy~\cite{Bulgac:2022,Bulgac:2023,Bulgac:2024a,Robin:2026}.

Our calculations show that the norm and Hamiltonian overlaps are not ever close in shape to a Gaussian and 
the Gaussian overlap approximation cannot be used in eGCM, see also Ref.~\cite{Kafker:2025a}.  Related issues concerning the agreement 
between exact TDGCM and TDGCM+GOA have been discussed in the literature, which show significant discrepancies 
between the two approaches~\cite{Verriere:2017}.

%%%%%%%%%%%%%%%%%%%%%%%%%%%%%%%%%%%%%%%%%%%%%%%%%%  
\begin{figure}[h]
\includegraphics[width=0.9\columnwidth]{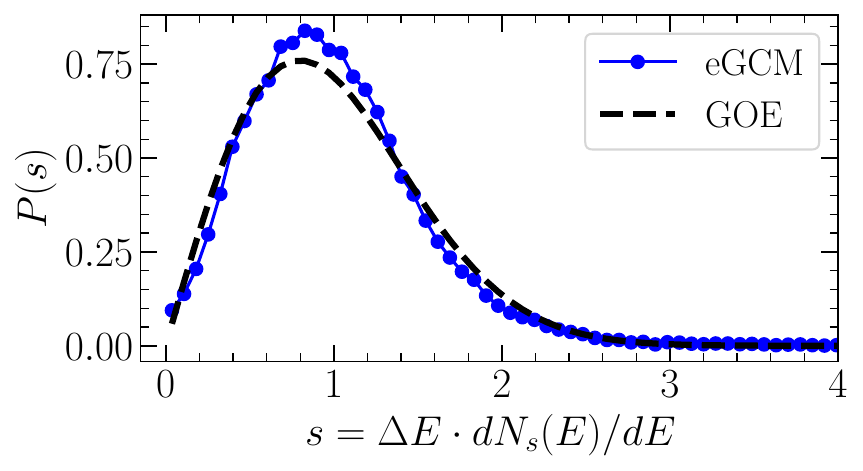}
\caption{\label{fig:6} The GOE level spacing distribution $P(s)$ compared to the corresponding unfolded eGCM spacing distribution. $N_s(E)$ stands for the unfolded cumulative density of states. The eGCM spectrum has also  a typical GOE oscillatory behavior  $\lim_{n\rightarrow \infty} |n-\sum_1^n s_k|^2  = {\cal O}(1)$~\cite{Bohigas:1993,Mehta:1991}. }
\end{figure}  
%%%%%%%%%%%%%%%%%%%%%%%%%%%%%%%%%%%%%%%%%%%%%%%%%  

%%%%%%%%%%%%%%%%%%%%%%%%%%%%%%%%%%%%%%%%%%%%%%%%%  
\begin{figure}[h]
\includegraphics[width=0.9\columnwidth]{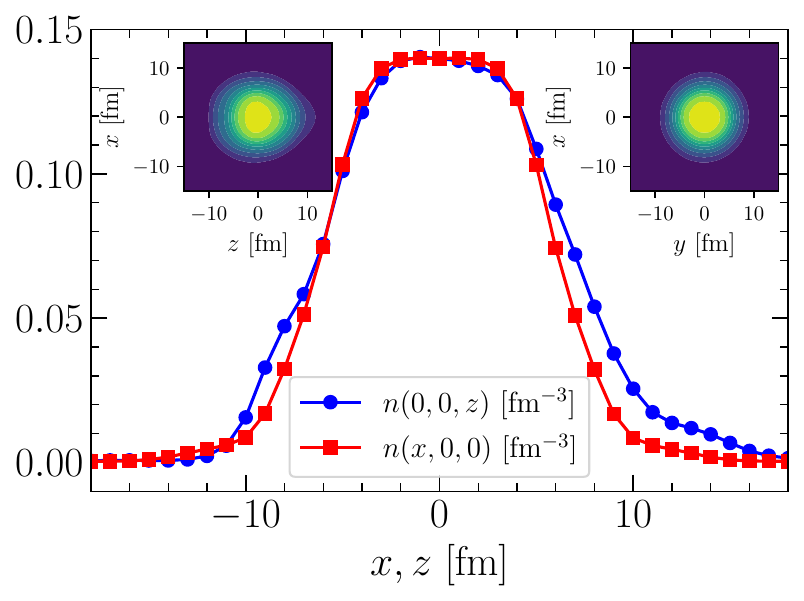}
\caption{\label{fig:7} The total nucleon density profiles of the emerged CN with 2D density profiles in the insets. The emerged CN average atomic number is $\langle A\rangle \approx 245.3$.}
\end{figure}  
%%%%%%%%%%%%%%%%%%%%%%%%%%%%%%%%%%%%%%%%%%%%%%%%%  

The static eGCM 
spectrum in Fig.~\ref{fig:eGCM_Energies} is ``almost degenerate and linear'' over most of the energy interval, 
with the comparison of level spacing distribution shown in Fig.~\ref{fig:6}. These are the states 
which define the structure of the initial eGCM wave function $\Psi(0)$, see Eqs.~(\ref{eq:psi_t}, \ref{eq:c_n}), 
when the colliding $^{48}$Ca and $^{208}$Pb are initially infinitely separated  
\begin{align}
&\!\!\! \!\!\! |\Psi(0)\rangle = {\cal A} \left ( e^{i{\bf K}\cdot {\bf R}} |\Phi_{^{48}Ca}\rangle \otimes |\Phi_{^{208}Pb}\rangle \right )\!= \!\sum_n c_n(0)|\Psi_n\rangle,\label{eq:ini_wf}
\end{align}  
where ${\cal A}$ stands for anti-symmetrization and appropriate normalization, ${\bf R}$ is the separation along 
the collision direction between the two nuclei, ${\bf K}$ the relative momentum of their initial center-of-mass energy at infinity, and 
with $c_n(0)$ defined in Eq.~\eqref{eq:c_n}. The coefficients $c_n$ for eigenvalues $E_n$ outside the energy interval (250, 36,500) MeV  are negligible ($\leq 10^{-16}$). 
As we discussed in the text, we considered only grazing collisions with impact parameters $b \in  [5, 6]$ fm, which are typically considered in MNT reactions~\cite{Simenel:2010,Sekizawa:2013,Sekizawa:2016,Simenel:2025,Simenel:2025a,Wu:2026}. For smaller impact parameters the two colliding nuclei 
do not separate in the time interval considered here in TDHF MNT reactions. 

The nucleon density profiles in Fig.~\ref{fig:7} and  the values $\langle r^2\rangle, \, Q_{20}, \, Q_{30}$ (see main text) appear to indicate
that full CN thermalization and ``memory loss'' has not been achieved at $t= 2,482$ fm/c.

\end{document}